\documentclass[prd,twocolumn]{revtex4}

\usepackage{graphicx, epsfig}
\usepackage{epstopdf}
\usepackage{color}
\usepackage{mathrsfs}
\usepackage{amsmath}

\newcommand{\be}{\begin{equation}}
\newcommand{\ee}{\end{equation}}
\newcommand{\bea}{\begin{eqnarray}}
\newcommand{\eea}{\end{eqnarray}}

\newcommand{\gapp}{\mathrel{\raise.3ex\hbox{$>$}\mkern-14mu
              \lower0.6ex\hbox{$\sim$}}}
\newcommand{\lapp}{\mathrel{\raise.3ex\hbox{$<$}\mkern-14mu
              \lower0.6ex\hbox{$\sim$}}}

\begin{document}
\title{Radiation from a collapsing object is manifestly unitary}
\author{Anshul Saini, Dejan Stojkovic}
\affiliation{ HEPCOS, Department of Physics, SUNY at Buffalo, Buffalo, NY 14260-1500, USA}


\begin{abstract}
The process of gravitational collapse excites the fields propagating in the background geometry and gives rise to thermal radiation.  We demonstrate by explicit calculations that the density matrix corresponding to such radiation actually describes a pure state. While Hawking's leading order density matrix contains only the diagonal terms, we calculate the off-diagonal
correlation terms. These correlations start very small, but then grow in time. The cumulative effect is that the correlations become comparable to the leading order terms and significantly modify the
density matrix. While the trace of the Hawking's density matrix squared decreases during the evolution,  the trace of the total density matrix squared remains unity at all times and all frequencies.
This implies that the process of radiation from a collapsing object is unitary.
\end{abstract}


\maketitle

{\it Introduction.~}
One of the most pressing problems in modern physics is the information loss paradox in black hole physics. Since Hawking radiation is purely thermal \cite{Hawking:1974sw}, it is possible to convert a pure state into a mixed state, which is forbidden in unitary quantum mechanics \cite{Hawking:1976ra}. It was often argued that subtle correlations between the emitted Hawking quanta which are usually neglected could be enough to recover information about the initial state and convert an apparently maximally mixed thermal state into a pure state \cite{Page:1979tc,Kraus:1994by}. This point of view was also often criticized by noticing that small corrections to the leading order Hawking terms are not enough to recover unitarity \cite{Mathur:2009hf}. Most of this discussion is given for the case of radiation from the pre-existing black hole, or in the limit when the horizon is already formed in infinite future. The purpose of this paper is to address this problem from a point of view of an asymptotic observer who is observing a time-dependent gravitational collapse. We find indeed that the process of gravitational collapse and subsequent evaporation is manifestly unitary as described in the foliation of an asymptotic observer.

We used the functional Schrodinger formalism which is especially convenient for this question since it gives us the time evolution of the system rather than radiation from a pre-existing black hole \cite{Vachaspati:2006ki,Vachaspati:2007hr,Greenwood:2008zg,Greenwood:2010mr,Greenwood:2008ht,Wang:2009ay,Halstead:2011gs,
Greenwood:2010uy,Greenwood:2009pd,Greenwood:2009gp,Greenwood:2010sx,Vachaspati:2007ur,Greenwood:2014bfa}. We start with a massive shell which is collapsing under its own gravitational pull. This process induces a non-trivial time-dependent metric which then excites the field quanta. The process of the gravitational collapse takes infinite time for an outside observer, however, radiation is pretty close to thermal when the collapsing shell approaches its own Schwarzschild radius. Our formalism gives us an explicit form of the wavefunction of the emitted radiation, which contains complete information not only about the diagonal Hawking terms, but also about the non-diagonal  correlations terms. Correlations between the Hawking quanta are at first indeed negligible with respect to the diagonal terms. However, time evolution creates progressively more off-diagonal terms than the diagonal ones. Moreover, time evolution is such that these cross-terms become of the same order of magnitude as the Hawking terms. As a result, the density matrix for the emitted radiation is significantly modified, in particular it is not purely diagonal.
We calculate the time evolution of the complete  density matrix as a function of time and frequency. The relevant quantity that we want to obtain is the trace of the density matrix squared ($Tr(\hat{\rho}^2)$), which tells us whether the system is in a pure or mixed state. We find that if we take only diagonal terms in density matrix then $Tr(\hat{\rho_h}^2)$ diverges from unity, which means that the state goes from pure to mixed. This is the standard Hawking's result which implies information loss. However if we include the off-diagonal terms then $Tr(\hat{\rho}^2)$ remains unity at all frequencies and all times during the evolution. This means that the initial state stays pure during the evolution. An observer who could measure the whole density matrix would not lose any information. This is the main result of our analysis.

{\it The formalism.~}
We consider a thin shell of matter which collapses under its own gravity. We use Schwarzschild coordinates because we are interested in the point of view of an observer at infinity. The metric outside the shell can be written as
\be
{ds}^2 =  - \left(1- \frac{R_s}{r}\right) dt^2 +  {\left(1- \frac{R_s}{r}\right)}^{-1} dr^2 + r^2 d{\Omega}^{2}
\ee
The interior of the shell is a flat spacetime due to the Birkhoff theorem
\be
{ds}^2 =  - dT^{2}  +  dr^2 + r^2 d{\Omega}^{2} .
\ee
The time coordinates of the two regions are related with the proper time inside the shell as
\be
\frac{dT}{d\tau}=\sqrt{1+{R_\tau}^2} \ ,
\label{dttau}
 \ \ \ \ \
\frac{dt}{d\tau} = \frac{ \sqrt{B + {R_\tau}^2}}{B}
\ee
where $B = 1 - R_s/R$ and $R_\tau = \frac{dR}{d\tau}$. From here we get
\be
\frac{dT}{dt} =  \sqrt{B - \left(\frac{1-B}{B}\right) {R_t}^2}
\label{dTdt}
\ee
An action of the massless scalar field propagating in the background of the collapsing shell can be written as
\be
S =  \int{ d ^{4} x \sqrt{-g} \frac{1}{2} g^{\mu \nu } {\partial}_{\mu} \phi  {\partial}_{\nu} \phi}
\ee
where $\phi$ is a scalar field, which we can expand in terms of the modes as
\be
\phi  = \sum_{\lambda} a_{\lambda} (t) f_{\lambda} (r) .
\ee
In the interior of shell,  the action takes the form
\be
S_{in} = 2 \pi \int{dt} \int_{0}^{R(t)} dr r^2 \left[ -  \frac{({\partial}_{t} \phi)^{2}}{ T_t} + T_t ({\partial}_{r} \phi)^{2} \right]
\ee
Similarly, outside of the shell it becomes
\be
S_{out} = 2 \pi \int{dt} \int_{R(t)}^{\infty} dr r^2 \left[ -  \frac{({\partial}_{t} \phi)^{2}}{ 1 - \frac{R_s}{r}} +\left(1 - \frac{R_s}{r}\right) ({\partial}_{r} \phi)^{2} \right]
\ee
The classical equation of motion for this collapsing shell  near the horizon can be written as \cite{Vachaspati:2006ki}
\be
R_t = \pm B \sqrt{1- \frac{B R^4}{h^2}}
\label{classol}
\ee
where $h$ is a constant. Using   Eq. (\ref{dTdt}) and Eq. (\ref{classol}), we get
\be \label{Tt}
T_t = B \sqrt{1 + (1-B)\frac{R^4}{h^2}}
\ee
When the shell is approaching its own Schwarzschild radius, $ R \rightarrow R_s$, then  $T_t \rightarrow 0$, hence the total action becomes
\bea
S \sim && 2\pi \int dt \left( - \frac{1}{B} \int_{0}^{R_s} dr r^2  ({\partial}_{t} \phi)^{2} + \right. \\
&& \left. \int_{R_s}^{\infty} dr r^2 \left(1 - \frac{R_s}{r}\right) ({\partial}_{r} \phi)^{2}\right) \nonumber
\eea
which in terms of the modes gives
\be \label{action}
S = \int dt \left( - \frac{1}{2B} \frac{d{a}_{k}}{dt} A_{kk\rq{}}\frac{d{a}_{k\rq{}}}{dt}  + \frac{1}{2}  a_{k} B_{kk\rq{}}a_{k\rq{}} \right)
\ee
with
\bea
&& A_{kk\rq{}} = 4 \pi \int_{0}^{R_s} dr r^2 f_k (r) f_{k\rq{}} (r)\\
&& B_{kk\rq{}} = 4 \pi \int_{R_s}^{\infty} dr r^2 \left( 1 - \frac{R_s}{r} \right) f\rq{}_k (r) f\rq{}_{k\rq{}} (r) .
\eea
Matrices $ A_{kk\rq{}}$ and $B_{kk\rq{}}$ are independent of R(t). From the action  (\ref{action}), we can find the corresponding Hamiltonian and write down the  Schrodinger equation $H \psi =  i \partial \psi / \partial t$ as
\be
\left[ \left( 1-\frac{R_s}{R}\right) \frac{1}{2} \Pi _k  {(A^{-1})}_{kk\rq{}} \Pi_{k\rq{}}  + \frac{1}{2} a_k B_{kk\rq{}} a_{k\rq{}} \right] \psi = i \frac{\partial \psi}{\partial t}
\ee
where the momentum is defined as
\be
\Pi_k = - i \frac{\partial}{\partial a_k} .
\ee
Since matrices $A$ and $B$ are symmetric and real, the principal axis theorem implies that both can be diagonalized simultaneously with respective eigenvalues $\alpha$ and $\beta$. One can then write the Schrodinger equation in terms of eigenmodes $y$ (which are linear combinations of the original modes $a$) as
\be
\left[ - \left(1-\frac{R_s}{R}\right)\frac{1}{2\alpha} \frac{{\partial}^2}{\partial y^2} + \frac{1}{2} \beta  y^2 \right] \psi(y, t) = i \frac{\partial \psi (y,t)}{\partial t} .
\label{schrodinger}
\ee
 Defining
\be \label{eta}
\eta = \int_{0}^{t} dt \left( 1 - \frac{R_s}{R}\right)
\ee
one can rewrite Eq.(\ref{schrodinger}) in form similar to the harmonic oscillator equation  as
\be
\left[ - \frac{1}{2\alpha} \frac{{\partial}^2}{\partial y^2} + \frac{\alpha}{2} {\omega}^2 (\eta) y^2 \right] \psi(y, \eta) = i \frac{\partial \psi (y,\eta)}{\partial \eta}
\ee
where
\be\label{baromega}
{\omega}^2(\eta) =\left(\frac{\beta}{\alpha}\right) \frac{1}{B} \equiv \frac{{\omega_0}^2}{B} .
\ee
Comparing Eq.~(\ref{eta}) with Eq.(\ref{Tt}), it is easy to see that $\eta$ is basically the time coordinate $T$ for an observer inside the shell in the near horizon limit.
The exact solution to this equation is \cite{Vachaspati:2006ki}
\be
\psi (y, \eta) = e^{i \delta (\eta)}{ \left[\frac{\alpha}{\pi  \theta^2}\right]}^{\frac{1}{4}} \exp \left[  i \left( \frac{\theta_\eta}{\theta} +\frac{i}{\theta^2}\right) \frac{\alpha y^2}{2} \right]
\ee
where  $\theta$ is the solution of the differential equation
\be
\theta_{\eta \eta} + {\omega}^2 (\eta)  \theta = \frac{1}{\theta^3}
\ee
with initial conditions
\bea
\theta(0) = \frac{1}{\sqrt{\omega_0}},   \theta_\eta (0) = 0 .
\eea
Since the background spacetime is time dependent, we make a distinction between the initial frequency $\omega_0$ at which the mode is created from the vacuum, and the final frequency at some later time $t$ defined as
\be
\bar{\omega} = \omega_0 e^{t/2} ,
\ee
where we used Eq.~(\ref{baromega}) and fact that from Eq.~(\ref{classol}) in the near horizon limit we have $B=e^{-t}$, where the time $t$ is expressed in units of $R_S$. However this frequency $\bar{\omega}$ is defined for the time parameter $\eta$, so to convert in to the the frequency defined for an asymptotic observer, $\Omega (t)$, we have to  perform a transform
\be
\Omega = \frac{d \eta}{dt} \bar{\omega} = e^{-t}\bar{\omega} = \omega_0 e^{-t/2} .
\ee

The wave function $\psi (y,t)$ contains information about the modes/particles excited in the spacetime in terms of their frequencies at the final moment $t$.
We want to construct density matrix of the system so we need to expand the wavefunction in terms of a complete basis. We will use the simple harmonic oscillator (SHM) basis  $\zeta_n (y)$.
\be
\psi (y,t) =  \sum_{n} c_n (t) \zeta_n (y)
\ee
The number of states in this basis is infinite so the size of the density matrix will be infinite too. However one can see that the probability of exciting higher $n$ states decreases rapidly as $n$ increases. Therefore one can easily identify trends even by considering finite (but large enough) $n$. The coefficients $c_n (t)$ can be written as
\be
c_n(t) =  \int dy {\zeta_n}^{*}(y) \psi (y,t) .
\ee
The probability of finding a particle in a particular state $n$ is given by ${\mid c_n(t)\mid}^2$ . The coefficients  $c_n$ can be explicitly found as (see supplemental material)
\begin{equation}
  c_n(t)=\frac{(-1)^{n/2}e^{i\alpha}}{(\Omega e^{t}\rho^2)^{1/4}}\sqrt{\frac{2}{P}}\left(1-\frac{2}{P}\right)^{n/2}\frac{(n-1)!!}{\sqrt{n!}}.
\end{equation}
where  $P$ is given by
\begin{equation}
  P=1-\frac{i}{\Omega e^{t}}\left(\frac{\theta_\eta}{\theta}+\frac{i}{\theta^2}\right).
\end{equation}

In order to find $c_n$ we need to solve for $\theta$. The simplest analytic method is given in \cite{Kolopanis:2013sty}.  $\theta$ and $\theta_\eta$ can be found in terms of  $\eta$ and $\xi$ as
\bea
\theta = \frac{1}{\sqrt{\omega_0}} \sqrt{\xi^2 + \chi^2 }.\\
\theta_\eta = \frac{1}{\omega_0\rho} (\xi \xi_\eta + \chi \chi_\eta ) .
\eea
$\eta$ and $\xi$ and their derivatives can be written in terms of Bessel's function as
\be
\xi = \frac{\pi u}{2} [ Y_0 (2\omega_0) J_1(u) - J_0(2\omega_0) Y_1(u)]
\label{xisol}
\ee
\be
\chi = \frac{\pi u}{2} [ Y_1 (2\omega_0) J_1(u) - J_1(2\omega_0) Y_1(u)]
\label{chisol}
\ee
\be
\xi_\eta = - \pi \omega_0^2 [ Y_0 (2\omega_0) J_0 (u) - J_0 (2\omega_0) Y_0(u)]
\label{xietasol}
\ee
\be
\chi_\eta = - \pi \omega_0^2 [ Y_1(2\omega_0) J_0(u) - J_1(2\omega_0) Y_0(u)]
\label{chietasol}
\ee
where $u \equiv 2\omega_0 \sqrt{1-\eta}$.

The occupation number at eigenfrequency $\Omega$  is given by the expectation value
\be \label{N}
N(t, \Omega) = \sum_n n |c_n|^2 .
\ee
The process of the gravitational collapse takes infinite time for an outside observer, however, radiation is pretty close to Planckian when the collapsing shell approaches its own Schwarzschild radius (see appendix).
Since we are already working in a near-horizon approximation, if we plot $N(t, \Omega)$ for some fixed late $t$, the spectrum will resemble the thermal Hawking distribution \cite{Vachaspati:2006ki}. [An important thing to notice here is that we obtain an approximately Planckian distribution though we did not trace out any degrees of freedom. When calculating radiation from a pre-existing horizon, one gets the Planckian distribution of emitted particles only when the infalling modes are traced out. In contrast, in the foliation of an asymptotic observer all the excited modes remain in the causal contact with an observer (even the modes inside the shell), and we are not allowed to trace out any modes.]
While the distribution (\ref{N}) keeps track only of the diagonal terms, we are here interested in correlations between the emitted quanta, which is contained not in the diagonal spectrum, but actually in the total density matrix for the system.

{\it Density Matrix.~}
Knowing the expansion coefficients $c_n$ explicitly, we can construct the density matrix.  The density matrix is defined as
\be
\hat{\rho}  = \sum \left|\psi\right>\left<\psi\right|
\ee
In our basis it can be re written as
\be \label{tdm}
\hat{\rho}  = \sum_{mn} c_{mn} \left|{\zeta}_{m}\right>\left<{\zeta}_{n}\right|
\ee
where $c_{mn} \equiv c_{m}c_{n}$. Original Hawking radiation density matrix, $\rho_h$, contains only the diagonal elements  $c_{nn}$, while the cross-terms   $c_{mn}$ for $m\neq n$ are absent. The off-diagonal terms represent interactions and correlations between the states. The rationale behind neglecting the cross-terms is that these correlations are usually higher order effects and will not affect the Hawking's result in the first order.   However, as argued recently in \cite{Hutchinson:2013kka}
(see also \cite{Papadodimas:2012aq}), during the process of Hawking radiation, the correlations may start off very small, but gradually grow as the process continues. It may happen at the end that these off-diagonal terms can modify the Hawking density matrix significantly enough to yield a pure sate.  The time-dependent functional Schrodinger formalism is especially convenient to test this proposal since it gives us the time evolution of the system.
\begin{figure}[htpb]
\begin{center}
       \includegraphics[height=0.30\textwidth,angle=0]{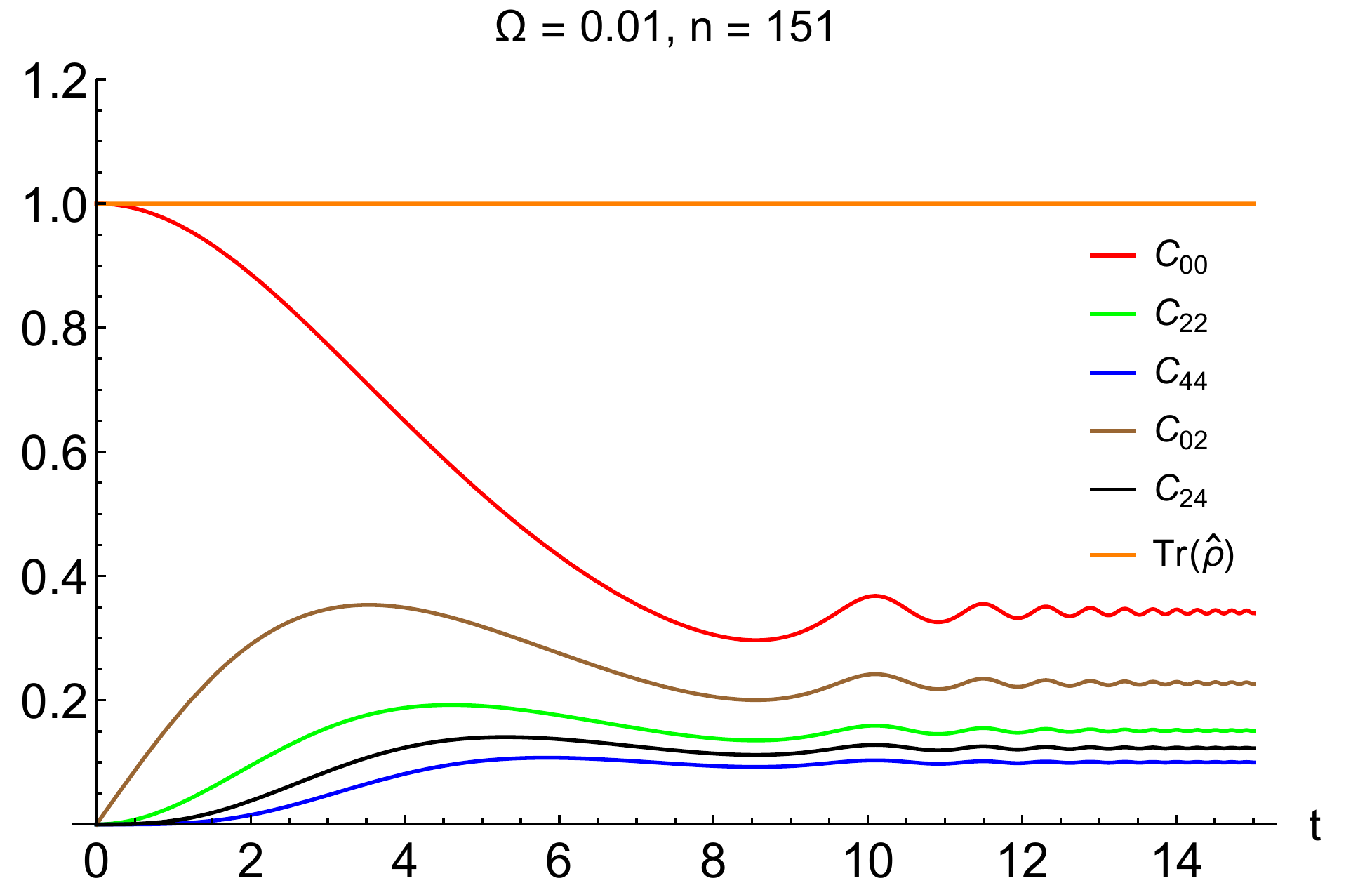}
 \caption{Elements of the density matrix  $c_{mn}$ and $Tr(\hat{\rho})$ as a function of time at $\Omega = 0.01$, where an index $n$ labeling the modes goes up to $n=151$. [Note that this is a final exponentially redshifted frequency, which was much higher at the time of the mode creation.] As time increases, the magnitude of $c_{00}$ decreases, $Tr(\hat{\rho})$ remains unity, and all other  $c_{mn}$ increase, reach the maximum values then decrease before reaching some form of a plateau. This implies that small correlations between the modes  become as important as the diagonal terms.}
 \label{cmn}
\end{center}
\end{figure}
In Fig.~\ref{cmn}, we plot some terms (both diagonal and off-diagonal) in the density matrix. We plot their time evolution with the fixed frequency  $\Omega$. We chose a small value of $\Omega$ to clearly demonstrate the effect (which is more pronounced for small frequencies). However we should note that $\Omega $ is an exponentially redshifted frequency, and the frequency of that mode at the time of creation was much higher.  We took absolute values of the off-diagonal $c_{mn}$ because they can be imaginary. All the units are dimensionless. Dimensionless frequency is given as $\Omega R_S$ , while dimensionless time is given as $t/R_S$.
From the plot one can see that the coefficient $c_{00}$ is initially almost exactly one, but then it decreases with time. The higher terms start small but then they increase with time, reach their maximum value and then they decrease before reaching some form of a plateau. This is expected because the system starts in the ground state. As time progress more modes are excited and higher order terms increase in magnitude. This increase of higher order terms can not proceed indefinitely if unitarity is preserved, i.e. any increase must be paid by a decrease somewhere else.
On the same plot, we show the trace of the density matrix  $Tr(\hat{\rho})$ as a check. The trace must remain unity at all times to preserve probabilities. However, we can numerically  take into account only a finite number of modes. Therefore, at some late time, the trace will start decreasing on the graph since higher modes which have not been included in numerics  will become important. The more modes we include, the longer the trace will remain unity.
In the supplemental material we proved that if one takes $n \rightarrow \infty$, then $Tr(\hat{\rho})$ always remains unity. Hence we plotted the graph only up to the time when $Tr(\hat{\rho})$ remains one.

What is more important is that the magnitudes of the  off-diagonal terms also increase with time. This implies that correlations among the created particles increase with time up to the point when even higher orders terms start increasing. Since there are progressively  more cross-terms than the diagonal terms, their cumulative contribution to the total density matrix simply can not be neglected.
\begin{figure}[htpb]
\begin{center}
  \includegraphics[height=0.30\textwidth,angle=0]{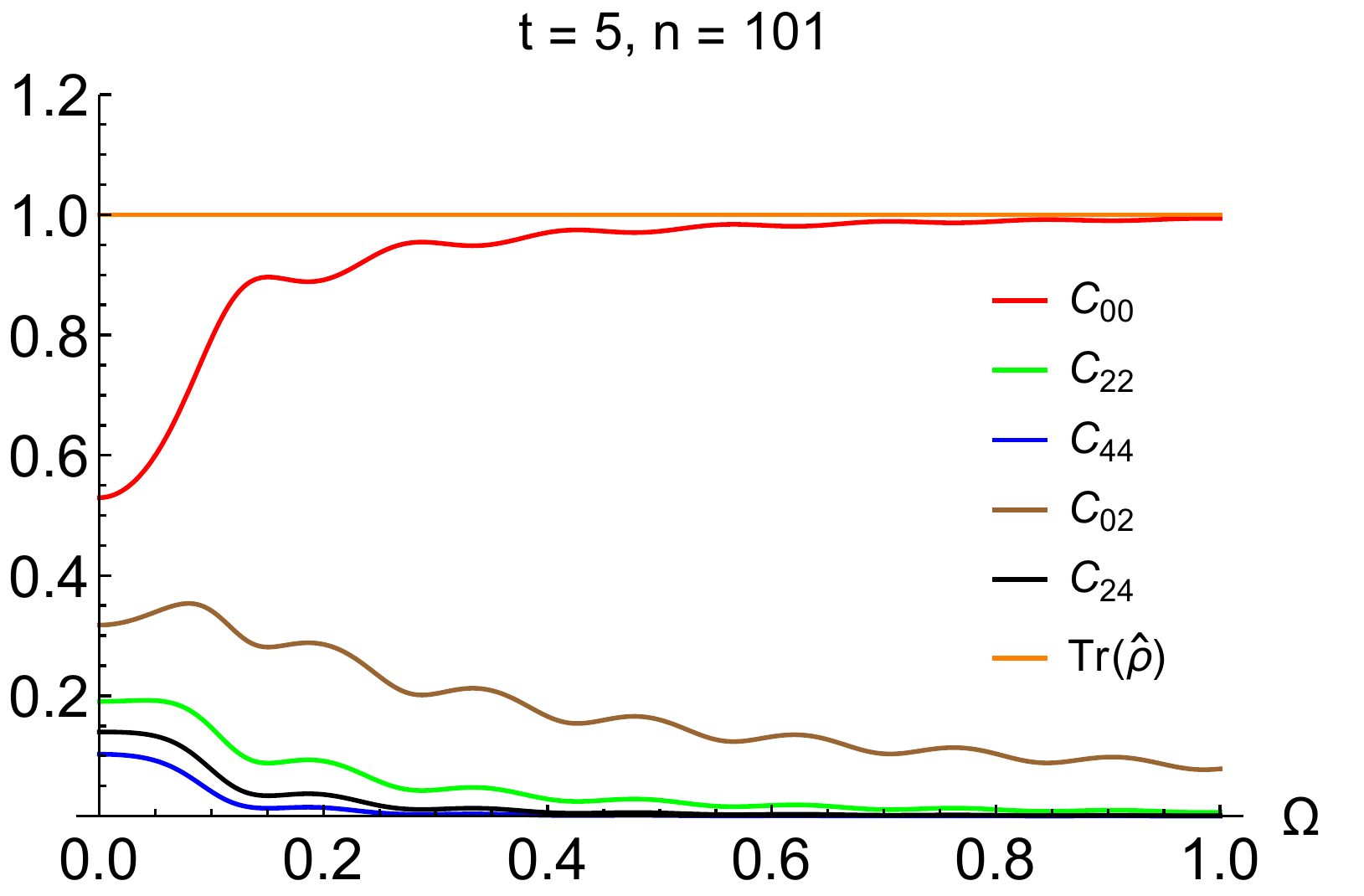}
\caption{
Cross-terms $c_{nm}$ and $Tr(\hat{\rho})$ as a function of $\Omega$ at fixed time $t = 5$. As $\Omega$ increases all $c_{mn}$  decrease, but $c_{00}$  increases. }
\label{cmnt}
\end{center}
\end{figure}
In Fig.~\ref{cmnt}, we plotted $c_{mn}$ and  $Tr(\hat{\rho})$ as a function of  $\Omega$ at a constant time. $Tr(\hat{\rho})$ remains one for all frequencies. The lowest term $c_{00}$ increases with $\Omega$, but all other terms decrease. This means that the
lowest diagonal term dominates and correlations are not that important at high frequencies.
\begin{figure}[htpb]
\begin{center}
  \includegraphics[height=0.30\textwidth,angle=0]{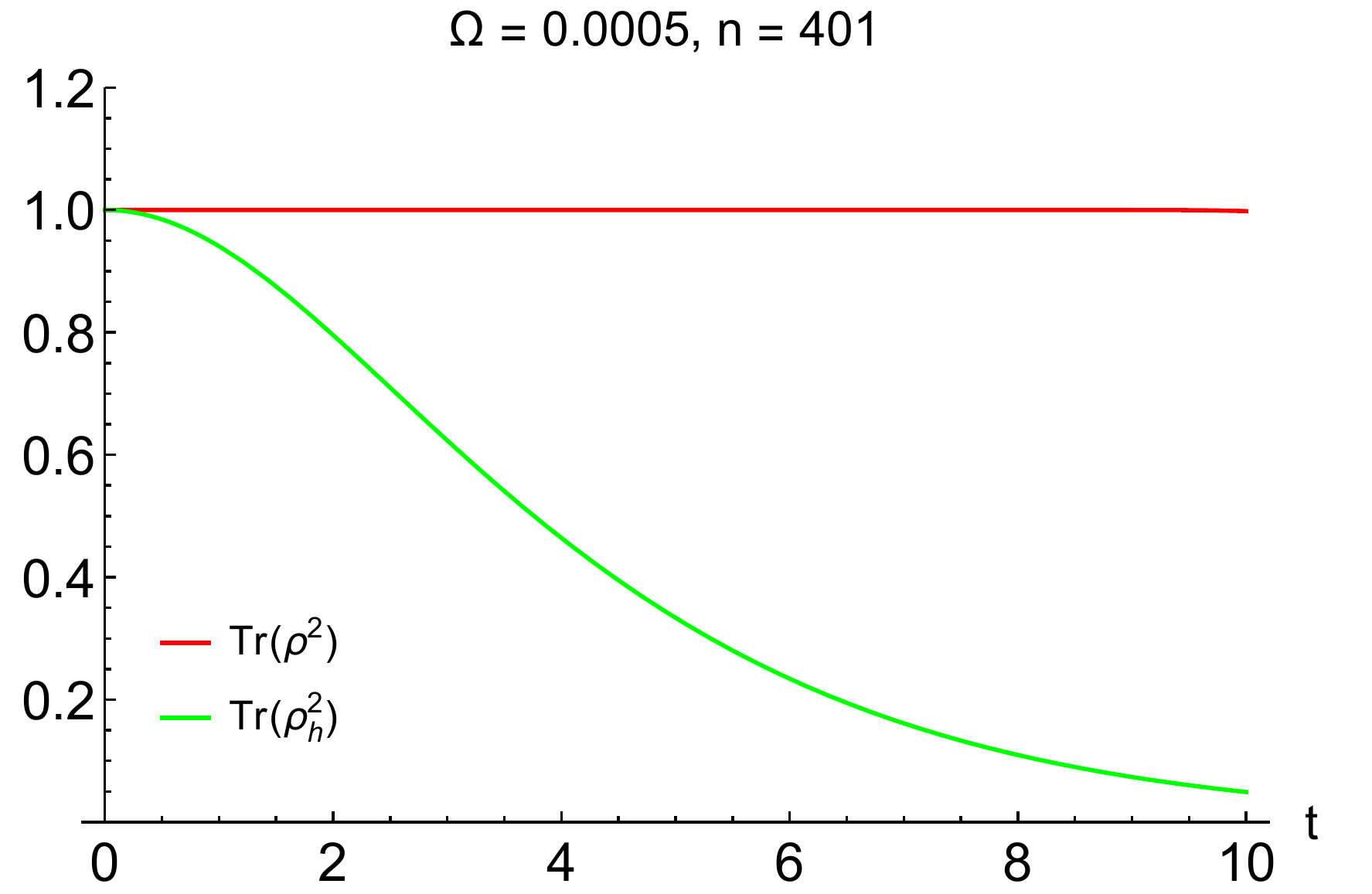}
 \caption{$\hat{\rho}_{h}$ is the diagonal Hawking density matrix, $\rho$ is the total density matrix as in Eq.~(\ref{tdm}). We plot $Tr(\hat{\rho}^2)$ and  $Tr(\hat{\rho}_h^2)$ as functions of time at a fixed final frequency $\Omega = 0.0005$.
The magnitude of  $Tr(\hat{\rho}_h^2)$ decreases with time which means that the system is losing information by going from a pure to a mixed state. However $Tr(\hat{\rho}^2)$
 remains unity at all times, which means that the state remains pure. This implies that the information of the system is conserved if cross-correlations are accounted for.  }
 \label{Trw}
\end{center}
\end{figure}
Information content in the system is usually given in terms of a trace of the squared density matrix. If the trace of the squared density matrix is one, then the state is pure, while the zero trace corresponds to a maximally mixed state. In Fig.~\ref{Trw}, we plot the traces of squares of two density matrices as functions of time for a fixed frequency. One is the Hawking density matrix $\hat{\rho}_{h}$ which contains only the diagonal terms $c_{nn}$ and neglects correlations. The other one is the total density matrix   $\hat{\rho}$ defined in Eq.~(\ref{tdm}) which contains all the elements, including the off-diagonal correlations. As expected,   $Tr(\hat{\rho}_h^2)$  decreases as time progress which means that the system is going from a pure state to a mixed thermal state. This is often labeled as the information loss in the process of Hawking radiation. However, if the plot the total $Tr(\hat{\rho}^2)$  we see that it always remains unity, which means that the state always remain pure during the evolution and information does not get lost. This clearly tells us that correlations between the excited modes are very important, and if one takes them into account the information in the system remains intact.
\begin{figure}[htpb]
\begin{center}
  \includegraphics[height=0.30\textwidth,angle=0]{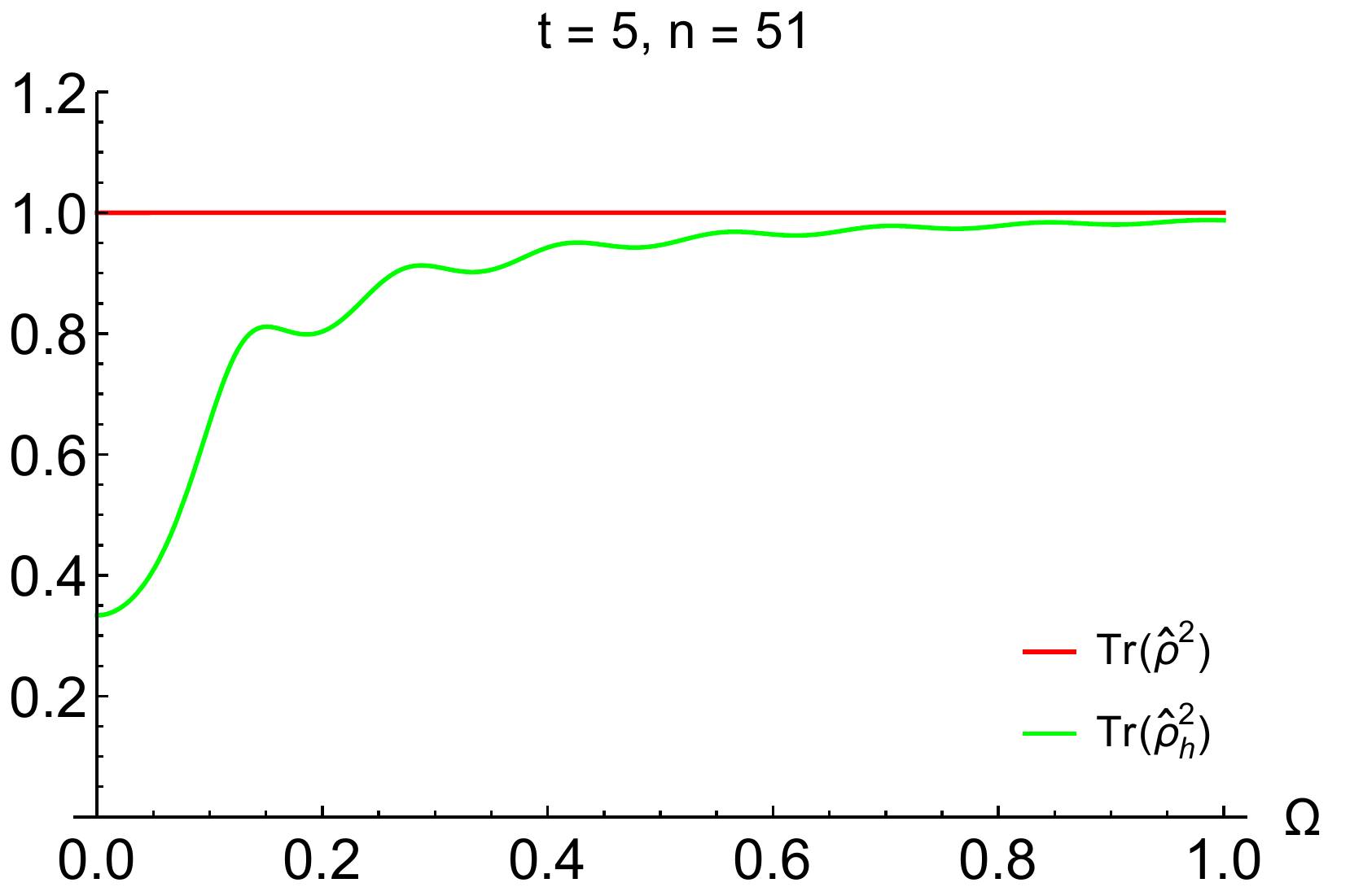}
\caption{ $Tr(\hat{\rho}^2)$ and  $Tr(\hat{\rho}_h^2)$ as a function of  $\Omega$ at $ t =5$. Again  $Tr(\hat{\rho}^2)$
remains one at all frequencies but  $Tr(\hat{\rho}_h^2)$ differs from unity at low frequencies which means it does not account for the full information of the system at low frequencies.  }
\label{Trt}
\end{center}
\end{figure}
In Fig.~\ref{Trt}, we plot $Tr(\hat{\rho}^2)$ and  $Tr(\hat{\rho}_h^2)$ as a function of  $\Omega$ at a fixed time. As expected,  $Tr(\hat{\rho}^2)$
remains one at all frequencies, but  $Tr(\hat{\rho}_h^2)$ differs from unity at low frequencies. This implies that $\rho_{h}$ gives a good description of the system at high frequencies,  but it fails to do so at low frequencies.

{\it Conclusions.~}
In conclusions, we showed by explicit calculations that radiation coming from a collapsing object is manifestly unitary in a space-time foliation that is motivated by an asymptotic observer. Hawking's thermal density matrix is diagonal and inevitably leads to information loss. However, when we take the off-diagonal correlation terms into account, the density matrix describes a pure state at all times.  This result agrees well with \cite{Page:1993df}, where it was shown at that at late enough time all the information in the system is contained in  correlations between the small subsystems (in this case emitted particles). For unitarity to be manifest to an observer, he has to be able to observe the total density matrix, i.e. all the created modes and correlations between them. In the foliation of an asymptotic observer all the excited modes remain in the causal contact with an observer (even the modes inside the collapsing shell since the collapsing object must disappear in finite time as seen by this observer), and we are not allowed to trace out any modes. This is in contrast with the usual procedure when Hawking radiation is calculated from a pre-existing horizon or in the limit when the horizon is formed at infinite future where the ingoing modes become causally separated from the outside observer.

If some of the modes are lost into the singularity (e.g. in the standard calculations like in \cite{Hawking:1976ra}), then the incomplete density matrix may not look like that of the pure state. However, there are strong indications that quantum effects should be able to rid gravity
of singularities, just as it was the case with the singular Coulomb potential (see for example \cite{Saini:2014qpa}).

Finally, our analysis was done for a static outside observer, however, it will be very important to learn what an infalling observer would see during the collapse in order to settle down the question of information loss.

\vskip.2cm {\bf Acknowledgment} \vskip.2cm
This work was also partially supported by the US National Science Foundation, under Grant No. PHY-1066278 and PHY-1417317.


\appendix

\section{Number of particles produced as a function of time~}

We work in a space-time foliation of an asymptotic observer. This observer measures frequency $\Omega$ according to his time parameter, $t$. From the original Schrodinger equation (Eq.17) expressed in the asymptotic time coordinate, $t$, we see that the mass term of the harmonic oscillator is $\alpha(t)$ which is related to a constant $\alpha$ as
\be
\alpha(t) = \frac{\alpha}{(1-\frac{R_s}{R})} = \alpha e^{t}.
\ee
To go to Eq.(19), we divide (Eq.17) with the factor $B=1 -R_s/R$. From Eq.(9) we have $B=e^{-t}$, where the time $t$ is expressed in units of $R_S$. This will rescale our time parameter from $t$ to $\eta$. We then proceed by solving Eq.(19), and the solution is given by Eq.(21). A solution to Eq.(19) must also be a solution to Eq.(17) with an appropriate rescaling.

To study the particle content in the system, we need to expand the wavefunction in Eq.(19) in terms of a complete basis. We use the simple harmonic oscillator (SHM) basis  $\zeta_n (y)$.
\be
\psi (y,t) =  \sum_{n} c_n (t) \zeta_n (y)
\ee
The coefficients $c_n (t)$ can be written as the overlap between the wavefunction and the basis states
\be
c_n(t) =  \int dy {\zeta_n}^{*}(y) \psi (y,t) .
\ee
The simple harmonic oscillator basis functions in terms of the frequency $\Omega$ measured by an asymptotic observer are
\begin{equation}
  \zeta_n (y)=\left(\frac{\alpha(t)\Omega}{\pi}\right)^{1/4}\frac{e^{-\alpha(t)\Omega y^2/2}}{\sqrt{2^nn!}}H_n(\sqrt{\alpha(t) \Omega}y)
\end{equation}
where $H_n$ are the Hermite polynomials.  A mode is created from vacuum at its initial frequency $\omega_0 = \sqrt{\frac{\beta}{\alpha}}$ (as in Eq.~(20)).
So in vacuum, $\Omega=\omega_0$. Since the background is time-dependent, this frequency will evolve in time according to Eq.(25).
Then, Eq.(21) together with Eq.(26) gives
\begin{align}
  c_n=&\left(\frac{1}{\Omega e^{t} \pi^2\theta^2}\right)^{1/4}\frac{e^{i\delta}}{\sqrt{2^nn!}}\int dx e^{-Px^2/2}H_n(x)\nonumber\\
           \equiv&\left(\frac{1}{\Omega e^{t}\pi^2\theta^2}\right)^{1/4}\frac{e^{i\delta}}{\sqrt{2^nn!}}I_n
\end{align}
where
\begin{equation}
  P=1-\frac{i}{\Omega e^{t}}\left(\frac{\theta_\eta}{\theta}+\frac{i}{\theta^2}\right).
\end{equation}

To find $I_n$ consider the corresponding integral over the generating function for the Hermite polynomials
\begin{align}
  J(z)&=\int dx e^{-Px^2/2}e^{-z^2+2zx}\nonumber\\
         &=\sqrt{\frac{2\pi}{P}}e^{-z^2(1-2/P)}
\end{align}
Since
\begin{equation}
  e^{-z^2+2zx}=\sum_{n=0}^{\infty}\frac{z^n}{n!}H_n(x)
\end{equation}
\begin{equation}
  \int dx e^{-Px^2/2}H_n(x)=\frac{d^n}{dz^n}J(z)\Big{|}_{z=0}
\end{equation}
Therefore
\begin{equation}
  I_n=\sqrt{\frac{2\pi}{P}}\left(1-\frac{2}{P}\right)^{n/2}H_n(0).
\end{equation}
Since
\begin{equation}
  H_n(0)=(-1)^{n/2}\sqrt{2^nn!}\frac{(n-1)!!}{\sqrt{n!}}
\end{equation}
and $H_n(0)=0$ for odd $n$, we find the coefficient $c_n$ for even values of $n$,
\begin{equation}
  c_n=\frac{(-1)^{n/2}e^{i\delta}}{(\Omega  e^{t} \theta^2)^{1/4}}\sqrt{\frac{2}{P}}\left(1-\frac{2}{P}\right)^{n/2}\frac{(n-1)!!}{\sqrt{n!}}.
\end{equation}
For odd $n$, $c_n=0$.

\section{Planckian spectrum}

In his original calculations, Hawking used the Bogoliubov transformation between the initial (Minkowski) vacuum and final (Schwarzschild) vacuum at the end of the collapse. The vacuum mismatch gives the thermal spectrum of particles. In this picture, there is a negative energy flux toward the center of a black hole and positive energy flux toward infinity, and thus a black hole loses its mass. Note that the existence of the horizon is necessary for this, since it is not possible to have a macroscopic negative energy flux without the horizon (the fact that the time-like Killing vector becomes space-like within the horizon is responsible for this). Since an outside observer, which is the most relevant observer for the question of the information loss, never sees the formation of the horizon, it is not clear how this picture works for him.

Here we plot the distribution of produced particles $N(t, \Omega) = \sum_n n |c_n|^2$ for several different times, and compare it with the Planckian distribution $N_{\rm Planck} (\Omega) =1/(e^{\Omega/T}-1)$, where $T$ is the temperature.
\begin{figure}[htpb]
\begin{center}
 \includegraphics[height=0.30\textwidth,angle=0]{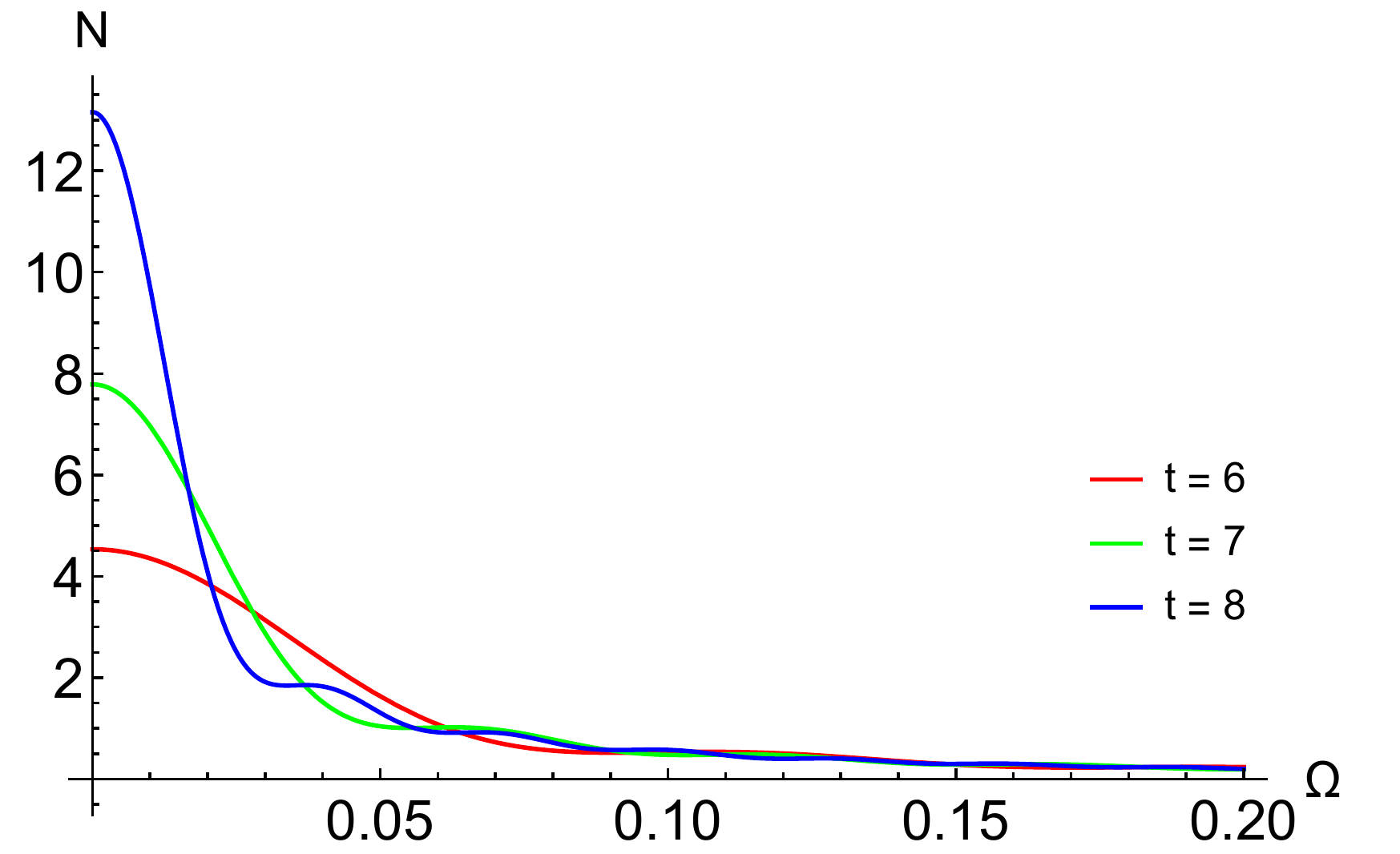}
\caption{Particle occupation number $N(\Omega)$ for several different times. As time increases, the distribution approaches the Planckian distribution $N_{\rm Planck}(\Omega)$ (in particular, $N(\Omega)$ becomes divergent at $\Omega \rightarrow 0$).}
\label{Nomega}
\end{center}
\end{figure}

This is where the advantage of the time dependent functional Schrodinger formalism becomes obvious. In this picture, it is the time dependent metric during the collapse which produces particles. No horizon is needed for this. As the collapsing shell approaches its own Schwarzschild radius, this radiation becomes more and more Planckian. Finally, at $t \rightarrow \infty$, when the horizon is formed it becomes completely Planckian and we recover Hawking’s result (his calculations were performed in $t \rightarrow \infty$).

Note that a static outside observer will never witness formation of the horizon since the collapsing object has only finite mass. An outside observer will observe the collapsing object slowly getting converted into Hawking radiation before horizon is formed. For him, no horizon nor singularity ever forms.
Still, if the radiation coming out from the collapsing object is thermal, and since there is a region inaccessible to an outside observer in Schwarzschild coordinates, this would lead to information loss. However, if radiation is not thermal, then information from the part of the space inaccessible to an outside observer could leak out.

We now emphasis that the Planckian spectrum of produced particles is not equivalent to a thermal spectrum. For a strictly thermal spectrum there should be no correlations between the produced particles. The corresponding density matrix should only have non-zero entries on its diagonal. In contrast, if subtle correlations exist, then the distribution might be Planckian, but the density matrix will have non-diagonal entries. If the state is pure, then the trace of the squared density matrix will be unity.

It is crucial for an observer to be able to see the whole density matrix, i.e. that he is able to measure all the modes and their correlations at least in principle.
In the foliation of an asymptotic observer that we worked in, all the excited modes remain in the causal contact with an observer (even the modes inside the collapsing shell), and we are not allowed to disregard any modes. In contrast, when calculating radiation from a pre-existing horizon, one has to trace out the infalling modes which eventually get lost into the singularity.  An important thing to notice here is that we obtained an approximately Planckian distribution though we did not trace out any degrees of freedom, which is again in contrast with calculations in the presence of the pre-existing horizon where thermal spectrum is obtained after tracing out the infalling modes.
Even in that case, thermality might be only apparent for some limited time. Singularity at the center is a classical result. Most likely, singularity at the center can be cured by quantization \cite{Saini:2014qpa}, just like we cured the atom of the classical $1/r$ singularity of the electrostatic potential. If there is no singularity at the center, and instead of it we find only a region of very strong but finite gravitational fields, then no modes are lost for an outside observer forever, and he can again measure the whole density matrix.

\section{Trace of the density matrix}

Here we will prove that the trace of the density matrix will add up to unity if $n \rightarrow \infty$. Let's take
\begin{equation}
  \xi=\left|1-\frac{2}{P}\right|.
\end{equation}
Now,

\begin{align}
  Tr(\hat{\rho})&=\sum_{n=even}\left|c_n\right|^2\nonumber\\
                                  &=\frac{2}{\sqrt{\Omega e^{t} \theta^2}|P|}\sum_{n=even}\frac{(n-1)!!}{n!!}\xi^n\nonumber\\
                                  &=\frac{2}{\sqrt{\Omega e^{t} \theta^2}|P|}\frac{1}{\sqrt{1-\xi^2}}\nonumber\\
                                  &=\frac{2}{\sqrt{\Omega  e^{t} \theta^2}|P|}\frac{1}{\sqrt{1-{\left|  1- \frac{2}{P}  \right|}^2}}
\end{align}
Inserting the expressions for $P$ and little algebra gives  $Tr(\hat{\rho}) = 1$.

\begin{figure}[htpb]
\begin{center}
 \includegraphics[height=0.30\textwidth,angle=0]{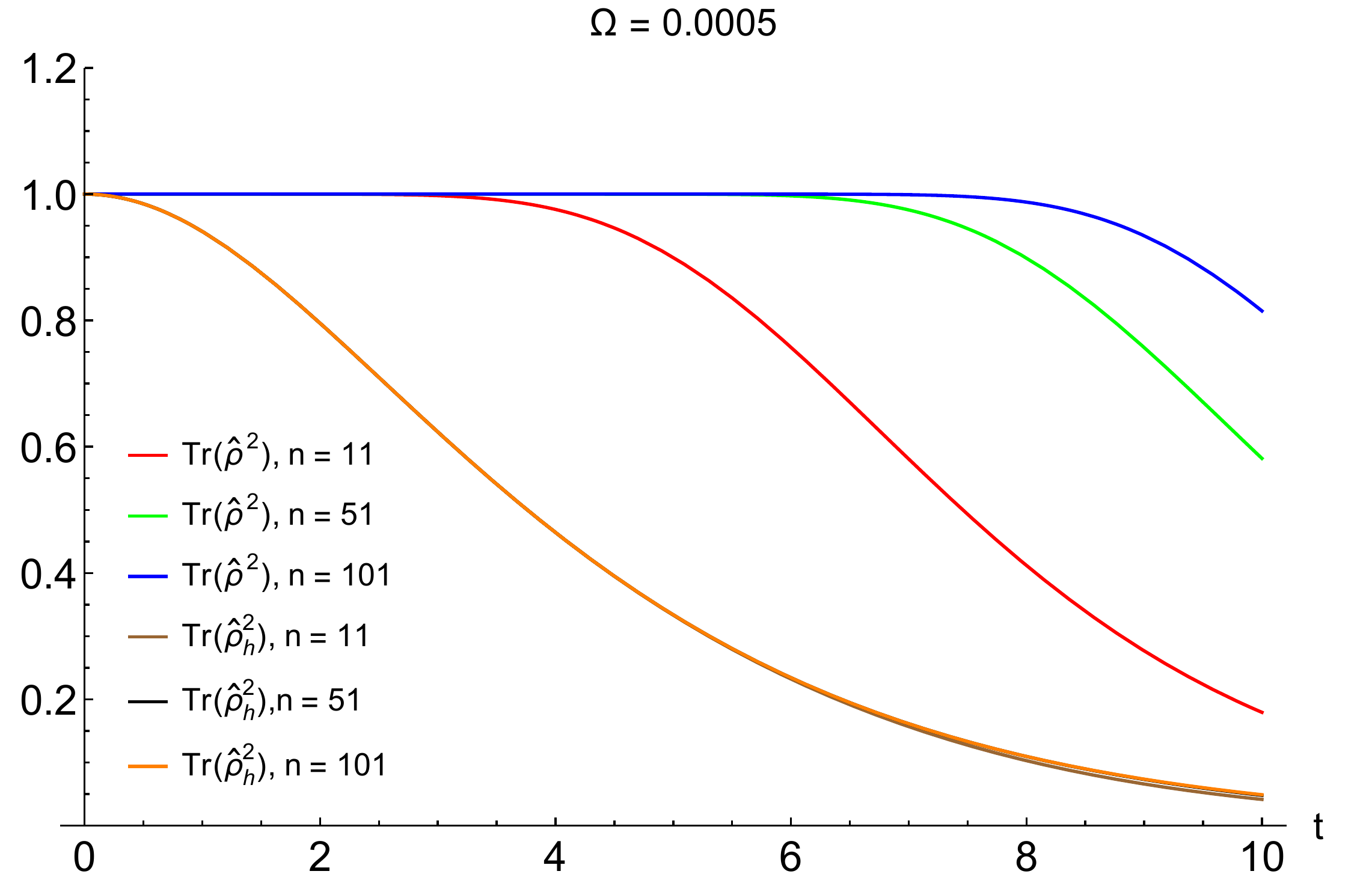}
\caption{ $Tr(\hat{\rho}_h^2)$ and $Tr(\hat{\rho}^2)$ are plotted for different values of $n$. $Tr(\hat{\rho}^2)$ remains unity as long as  $Tr(\hat{\rho})$ is unity. However,  the Hawking  $Tr(\hat{\rho}_h^2)$ drops from unity much earlier, and increasing $n$ does not affect it much (several curves with different $n$ lie on top of each other).}
\label{n}
\end{center}
\end{figure}

It is instructive to check also how the number of excited modes $n$ that we include in the calculations influences the behavior of  Tr($\hat{\rho}$), $Tr(\hat{\rho}_h^2)$ and $Tr(\hat{\rho}^2)$ with time. The trace of the original density matrix Tr($\hat{\rho}$) must remain unity at all times to preserve probabilities. However, we can numerically  take into account only a finite number of modes. Therefore, at some late time, the trace will start decreasing on the graph since higher modes which have not been included in numerics  will become important. The more modes we include, the longer the trace will remain unity.  We can also see in Fig.~\ref{n}  that $Tr(\hat{\rho}^2)$ remains unity as long as  $Tr(\hat{\rho})$ is unity. However,  the Hawking  $Tr(\hat{\rho}_h^2)$ drops from unity much earlier, and several curves with different $n$ lie on top of each other. Since $\hat{\rho}_h$ has only diagonal terms, this tells us that the off-diagonal correlation terms  indeed play crucial role in preserving purity of the state.

\end{document}